\documentclass[aps,pra,twocolumn,superscriptaddress,groupedaddress,footinbib,showpacs]{revtex4}
\usepackage[utf8]{inputenc}
\usepackage[english]{babel}
\usepackage{amsmath}
\usepackage{amsfonts}
\usepackage{amssymb}
\usepackage{amsbsy}
\usepackage{color,graphicx}
\usepackage{braket}
\usepackage{bbm,bm}
\usepackage{comment}
\usepackage[sort&compress]{natbib}

\usepackage[colorlinks=true,citecolor=blue,linkcolor=red,urlcolor=blue]{hyperref}

\newcommand{\bq}{\begin{equation}}
\newcommand{\eq}{\end{equation}}
\newcommand{\bqali}{\bq\begin{aligned}}
\newcommand{\eqali}{\end{aligned}\eq}
\newcommand{\bqn}{\begin{equation*}}
\newcommand{\eqn}{\end{equation*}}
\newcommand{\TR}[2]{\text{Tr}^{(\text{\tiny #1})}\left[#2\right]}
\newcommand{\Tr}[1]{\text{Tr}\left[#1\right]}
\newcommand{\Det}[1]{\text{det}\left[#1\right]}
\newcommand{\tran}{^{\mkern-1.5mu\mathsf{T}}}

\begin{document}
\title{Adjoint master equation for Quantum Brownian Motion}
\author{Matteo Carlesso}
\email{matteo.carlesso@ts.infn.it}
\affiliation{Department of Physics, University of Trieste, Strada Costiera 11, 34151 Trieste, Italy}
\affiliation{Istituto
Nazionale di Fisica Nucleare, Trieste Section, Via Valerio 2, 34127 Trieste,
Italy}
\author{Angelo Bassi}
\email{bassi@ts.infn.it}
\affiliation{Department of Physics, University of Trieste, Strada Costiera 11, 34151 Trieste, Italy}
\affiliation{Istituto
Nazionale di Fisica Nucleare, Trieste Section, Via Valerio 2, 34127 Trieste,
Italy}
\date{\today}

\begin{abstract}
Quantum brownian motion is a fundamental model for a proper understanding of open quantum systems in different contexts such as chemistry, condensed matter physics, bio-physics and opto-mechamics. In this paper we propose a novel approach to describe this model. We provide an exact and analytic equation for the time evolution of the operators, and we show that the corresponding equation for the states is equivalent to well-known results in literature. The dynamics is expressed in terms of the spectral density, regardless the strength of the coupling between the system and the bath. Our allows to compute the time evolution of physically relevant quantities in a much easier way than previous formulations allow to. An example is explicitly studied.
\end{abstract}
\maketitle

\section{Introduction}

Technical improvements in quantum experiments are making impressive
steps forward, reaching levels of accuracy which were hardly
imaginable a few decades ago.  Controlling the noise is often the
crucial challenge for further progress, and a theoretical
understanding is important to disentangle environmental effects from
intrinsic properties of the system.

Quantum Brownian motion \cite{Weiss:1999aa, Breuer:2002aa,
Joos:2003aa, Gardiner:2004aa, Schlosshauer:2007aa,Diosi:2014aa} is the paradigm of
an open quantum system interacting with an external bath, 
and nowadays it finds applications in several physical
contexts such as chemistry \cite{Pomyalov:2005}, condensed matter
\cite{Hope:1997aa,Breuer:2001aa,Jiang:2011}, bio-physics
\cite{Scholes:2011, Blankenship:2011aa, Fassioli:2012aa,
Lambert:2013aa} and opto-mechanics \cite{Reichel:2001aa,
Paternostro:2006aa, Hunger:2010aa, Groblacher:2015aa} to name a few.

The model consists of a particle $S$ of mass $M$,
with position $\hat x$ and momentum $\hat p$, harmonically trapped at frequency $\omega_\text{\tiny S}$ and interacting with a thermal
bath of independent harmonic oscillators, with positions $\hat
R_k$, momenta $\hat P_k$, mass $m$ and frequencies $\omega_k$.  This
model 
has become a milestone in the theory of open quantum system
\cite{Toda:1958aa,Magalinskii:1959aa,Senitzky:1960aa,Schwinger:1961aa,Feynman:1963aa,Ford:1965aa,Ullersma:1966aa, Caldeira:1983aa, Grabert:1984aa,Lindenberg:1984aa,Riseborough:1985aa,Haake:1985aa, Ford:1987aa,
Unruh:1989aa, Hu:1992aa, Breuer:2002aa,Ferialdi:2016ab}.  The total Hamiltonian $\hat
H_\text{\tiny T}$ of system plus bath is
$\hat H_\text{\tiny T}=\hat H_\text{\tiny S}+\hat H_\text{\tiny I}+\hat
H_\text{\tiny B},
$
 where
\bqali\label{001Eq.M.1}
\hat H_\text{\tiny S}&=\frac{\hat p^2}{2M}+\frac{1}{2}M\omega_\text{\tiny S}^2\hat
x^2,\\
 \hat
H_\text{\tiny B}&=\sum_k\frac{\hat P^2_k}{2m}+\frac{1}{2}m\omega_k^2\hat R_k^2,\\
\hat H_\text{\tiny I}&=\hat x\sum_kC_k\hat R_k,
\eqali
are respectively the system, bath and interaction Hamiltonians.
  The characterization of the set
of coupling constants $C_k$ is provided by the spectral density which
is defined as
\bq\label{001Eq.SE.10}
J(\omega)=\sum_k\frac{C_k^2}{2m\omega_k}\delta(\omega-\omega_k).
\eq
The first master equation for this model was derived by Caldeira and Leggett
\cite{Caldeira:1983aa} by using the common
assumption of a factorized initial state
\bq\label{001Eq.M.2}
\hat \rho_\text{\tiny T}(0)=\hat
\rho_\text{\tiny S}(0)\otimes\hat\rho_\text{\tiny B},
\eq
where $\hat\rho_\text{\tiny S}(0)$ and $\hat\rho_\text{\tiny B}$ are the initial states of the system and of the bath respectively, 
and also the Born-Markov approximation
\cite{Breuer:2002aa}; The high temperature limit was taken into
account in order to obtain a simple evolution
\cite{Kossakowski:1972aa,Gorini:1976aa,Lindblad:1976aa} in the Lindblad
form 
with constant coefficients:
\bqali\label{001Eq.M.1.3}
\dfrac{d\hat \rho_\text{\tiny S}(t)}{dt}=-\dfrac{i}{\hbar}[\hat H_\text{\tiny S},\hat \rho_\text{\tiny S}(t)]-\dfrac{i\gamma}{\hbar}\left[\hat x,\{\hat p,\hat \rho_\text{\tiny S}(t)	\}	\right]+\\
-\dfrac{2M\gamma}{\hbar^2\beta}\left[\hat x,[\hat x,\hat \rho_\text{\tiny S}(t)]	\right],
\eqali
where $\gamma$ and $\beta$ are the damping rate and the inverse temperature, respectively.
This derivation has two limitations. First,
the master equation is the 
generator of a dynamical map which is not positive
\cite{Ambegaokar:1991aa,Diosi:1993aa}, i.e.~it does not map all quantum
states $\hat \rho_\text{\tiny S}$ into quantum states \footnote{Although the Caldeira-Leggett master equation \eqref{001Eq.M.1.3} is not completely positive, it becomes of use when large time scales are considered. In fact, after a transient time of the order of $\sim\gamma^{-1}$, the dynamical map corresponding to Eq.~\eqref{001Eq.M.1.3} maps quantum states in quantum states. This is however still not sufficient to obtain in general the same asymptotic expectation values one obtains from the exact model. In order to avoid subsequent gross miscalculations, one needs to implement some effective modifications on the initial conditions \cite{Haake:1983aa,Geigenmuller:1983aa,Haake:1985aa}.}. Second, the regime of validity cannot be always fulfilled: the latest attempts 
to reach the ground state at low temperature regimes \cite{Groblacher:2015aa,Frimmer:2016aa}
is an opto-mechanical example.

The main contributions in overcoming these limitations were given by Haake and Reibold \cite{Haake:1985aa} and later by Hu, Paz and Zhang  \cite{Hu:1992aa}, who
provided the {exact} master equation for the particle $S$ given
the total Hamiltonian $\hat H_\text{\tiny T}$:
\begin{multline}\label{HPZmaster} \frac{d\hat
\rho_\text{\tiny S}(t)}{dt}=-\frac{i}{\hbar}[\hat H(t),\hat
\rho_\text{\tiny S}(t)]-\frac{i\gamma(t)}{\hbar}\left[\hat x,\{\hat p,\hat
\rho_\text{\tiny S}(t) \} \right]+\\ -h(t)\left[\hat x,[\hat x,\hat \rho_\text{\tiny S}(t)]
\right]-f(t)\left[\hat x,[\hat p,\hat \rho_\text{\tiny S}(t)] \right],
\end{multline}
where $\hat H(t)$ and the coefficients $\gamma(t)$, $h(t)$ and $f(t)$
now are time dependent. We refer to this model as to the Quantum Brownian Motion (QBM) model.
Contrary to the Caldeira-Leggett master equation, which is valid only for the
specific ohmic choice the spectral density ($J(\omega)\propto\omega$) Eq.~\eqref{HPZmaster} is valid for arbitrary spectral densities
$J(\omega)$ and temperatures $T$. However, for the QBM model, the coefficients are solutions of differential equations, which in general are hard to solve. 
 The explicit form of these coefficients, beyond the weak-coupling limit \cite{Hu:1992aa}, 
was provided by Ford and O'Connell in \cite{Ford:2001aa}.

The generality of such a solution is outstanding, however, as noticed
in \cite{Ford:2001aa}, solving the time-dependent master equation
 is in general a {formidable} problem. The authors show that the dynamics of 
 the system can be more easily solved by working with the Wigner function
 of the system and bath at time $t$ and then averaging over 
 the degrees of freedom of the bath. According to their procedure,
 the reduced Wigner function $W$ at time $t$ can be expressed in terms of
that at time $t=0$ as follows:
 \bq\label{Wignert}
 W(x,p,t)=\int_{-\infty}^{+\infty}dr\int_{-\infty}^{+\infty}dq P(x,p;r,q;t)W(r,q,0),
 \eq
 where $P$ describes the transition probability \cite{Ford:2001aa}. 
 
The drawback of such a procedure is the limited selection of initial states $\hat \rho_\text{\tiny S}$ for which
the Wigner function is analytically computable. For gaussian states 
this is not a problem, however there exist physical relevant situations where this is not the
case \cite{Ghose:2007aa,Gomez:2015aa,Huang:2016aa}. An example is provided by a system initially 
confined in a infinite square potential. We will refer explicitly to this example.

In this paper, we propose an alternative derivation of the QBM dynamics for a general bath at arbitrary temperatures. The master equation we derive is exact and of course is equivalent to Eq.~\eqref{HPZmaster}. However, the time-dependent coefficients will be written in a much simpler form, and therefore can be used to compute much more easily the solution of the master equation, regardless of the strength of the coupling and of the form of the initial state $\hat \rho_\text{\tiny S}$.\\

The paper is organized as follows:
In section \ref{sec.algebraic} we describe our alternative derivation. In section \ref{subsec:HPZ} we derive the master equation. In section \ref{sec.cp} we provide a criterion for the complete positivity of the dynamics. In section \ref{general_sol} we derive the explicit evolution of some physical quantities one is typically interested in, for a specific spectral density, in particular we compare our result with that of Haake-Reibold \cite{Haake:1985aa} and Hu-Paz-Zhang  \cite{Hu:1992aa}. Moreover, we will show how one can easily go beyond the results of Ford and O'Connell \cite{Ford:2001aa} and compute the time evolution of expectations values for initial non-gaussian states.

\section{The QBM model in the Heisenberg picture: the adjoint master equation}
\label{sec.algebraic}

We derive the adjoint master equation for the quantum Brownian motion model. This is the dynamical equation describing the time evolution of a generic operator $\hat O$ of the system $S$, once the average over the bath is taken. To this end, we consider the unitary time evolution of the extended operator $\hat O\otimes \bf{\hat1_B}$ with respect to the total Hamiltonian $\hat H_\text{\tiny T}$ of the system plus bath, where $\bf{\hat1_B}$ is the bath identity operator, and we trace over the degrees of freedom of the bath. The time derivative of the reduced operator, under the hypothesis of a factorized initial state as in Eq.~\eqref{001Eq.M.2}, will be governed by the adjoint master equation.

Let us consider the von Neumann representation \cite{Neumann:1931aa,Hiley:2015aa} of the operator $\hat O$, defined, at time $t=0$, by the following relation:
\bq\label{001Eq.H.0.4}
\hat O
=\int d\lambda d\mu \ \mathcal O(\lambda,\mu)\hat \chi(\lambda,\mu,t=0),
\eq
where $\mathcal O(\lambda,\mu)$ is the kernel of the operator $\hat O$ and $\hat \chi(\lambda,\mu,t=0)=\exp[i\lambda \hat x+i\mu \hat p]$ is the generator of the Weyl algebra, also called characteristic or Heisenberg-Weyl operator \cite{Hiley:2015aa}. Following the procedure previously outlined, and using the von Neumann representation, the operator $\hat O$ at time $t$ is given by:
\bq\label{001Eq.H.vonNt}
\begin{aligned}
\hat O_t=\int d\lambda d\mu \ \mathcal O(\lambda,\mu)\hat \chi_t,
\end{aligned}
\eq
where we introduced the characteristic operator at time $t$:
\bq
\hat \chi_t=\TR{B}{\hat \rho_\text{\tiny B}\left(\hat{\mathcal U}_t^{\dag}(\hat \chi(\lambda,\mu,0)\otimes {\bf\hat1_B})\hat{\mathcal U}_t\right)}.
\eq
and $\hat{\mathcal U}_t=\exp(-\tfrac i\hbar\hat H_\text{\tiny T}t)$. Therefore, to obtain the evolution of the operator $\hat O_t$, it is sufficient to consider the evolution equation for the characteristic operator:
\bq\label{001Eq.H.0.6}
\hat \chi_t=\TR{B}{\hat \rho_\text{\tiny B}
e^{i\lambda \hat x(t)+i\mu \hat p(t)}},
\eq
where $\hat x(t)$ and $\hat p(t)$ are the position and momentum operators of the system $S$ evolved by the unitary evolution generated by the total Hamiltonian of the composite system plus bath and $\hat \rho_\text{\tiny B}$ is defined in Eq.~\eqref{001Eq.M.2}.\\

In order to obtain the explicit expression of $\hat x(t)$ and $\hat p(t)$, we rewrite the bath and interaction Hamiltonian defined in Eq.~\eqref{001Eq.M.1} in terms of the creation and annihilation operators $\hat b^\dag_k$ and $\hat b_k$ of the $k$-th bath oscillator: $\hat H_\text{\tiny B}=\sum_k\hbar\omega_k\hat b_k^\dag\hat b_k$ and $\hat H_\text{\tiny I}=-\hat x\hat B(0)$, where $\hat B(t)$ is defined as 
\bq
\label{001Eq.H.4bis}
\hat B(t)=-\sum_kC_k\sqrt{\dfrac{\hbar}{2m\omega_k}}\left(\hat b_ke^{-i\omega_kt}+\hat b_k^{\dag}e^{i\omega_kt}\right).
\eq
In terms of the latter we solve the Heisenberg equations of motions for $\hat x(t)$ and $\hat p(t)$ by using the Laplace transform. The solutions are:
\begin{subequations}\label{001Eq.H.0.6b}
\begin{align}
\hat x(t)&=G_1(t)\hat x+G_2(t)\dfrac{\hat p}{M}+\dfrac{1}{M}\int_0^{t}ds\ G_2(t-s)\hat B(s),
\label{001Eq.H.0.6b.prima}\\
\hat p(t)&=M\dot G_1(t)\hat x+\dot G_2(t)\hat p+\int_0^{t}ds\ \dot G_2(t-s)\hat B(s),
\end{align}
\end{subequations}
where $\hat x$ and $\hat p$ denote the operators at time $t=0$, and the two Green functions $G_1(t)$ and $G_2(t)$ are defined as 
\bqali\label{001Eq.G.1.12}
G_1(t)&=\dfrac{d}{dt}G_2(t),\\
G_2(t)&=\mathcal L^{-1}\left[\dfrac{M}{M(s^2+\omega_\text{\tiny R}^2)- \mathcal L[ D(t)](s)/\hbar}\right](t),
\eqali
where $\mathcal L$ denotes the Laplace transform and $\omega_\text{\tiny R}^2=\omega_\text{\tiny S}^2+\tfrac{2}{M}\int d\omega\,J(\omega)/\omega$. In Eq.~\eqref{001Eq.G.1.12} we introduced the dissipation kernel $D(t)$:
\bq
\label{001Eq.SE.14a}
D(t)=2\hbar\int_0^{+\infty}d\omega\, J(\omega)\sin(\omega t).
\eq
Given Eqs.~\eqref{001Eq.H.0.6b}, since the operators of the system and of the bath commute at the initial time, it follows that:
\bq\label{001Eq.H.0.7.1}
\hat \chi_t=e^{i\alpha_1(t) \hat x+i\alpha_2(t) \hat p}\ \TR{B}{\hat \rho_\text{\tiny B}\hat \chi_\text{\tiny B}(t)},
\eq
where $\alpha_1(t)$ and $\alpha_2(t)$ are defined as follows:
\begin{subequations}\label{001Eq.H.0.7.bis}
\begin{align}
\label{001Eq.H.0.7.bis1}
\alpha_1(t)&=\lambda G_1(t)+\mu M\dot G_1(t),\\
\label{001Eq.H.0.7.bis2}
\alpha_2(t)&=\lambda G_2(t)/M+\mu \dot G_2(t),
\end{align}
\end{subequations}
and the operator $\hat \chi_\text{\tiny B}(t)$ refers only to the degrees of freedom of the bath:
\bq\label{eq:17}
\hat \chi_\text{\tiny B}(t)=\exp\left[i \int_0^t ds\ \hat B(s)\alpha_2(t-s)\right].
\eq
Under the assumption of a thermal state for the bath:
\bq\label{eq:18}
\hat \rho_\text{\tiny B}\propto e^{-\beta\hat H_\text{\tiny B}},
\eq
the trace over $\hat \chi_{\beta}(t)$ gives a real and positive function of time $\TR{B}{\hat \rho_\text{\tiny B}\hat \chi_\text{\tiny B}(t)}=e^{\phi(t)}$,
where the explicit form of $\phi(t)$ can be obtained by using the definition of the spectral density in Eq.~\eqref{001Eq.SE.10}. 
In Appendix \ref{app:A} we present the explicit form of $\phi(t)$, written as the sum of three terms: $\phi(t)=\lambda^2\phi_1(t)+\mu^2\phi_2(t)+\lambda\mu\phi_3(t)$.
The time derivative of $\hat \chi_t$ gives
\bqali\label{001Eq.H.0.7.c}
\dfrac{d\hat \chi_t}{d t}&=\Bigl[	i\dot\alpha_1(t)\hat x+i\dot\alpha_2(t)\hat p
+\Bigr.\\
&+\Bigl.\dfrac{i\hbar}{2}\bigl[	\dot\alpha_1(t)\alpha_2(t) - \alpha_1(t)\dot\alpha_2(t)\bigr]+\dot\phi(t)\Bigr]\hat \chi_t;
\eqali
by substituting this expression in:
\bq\label{001Eq.H.0.5.bis}
\dfrac{d}{dt}\hat O_t=\int d\lambda d\mu \ \mathcal O(\lambda,\mu)\dfrac{d\hat \chi_t}{d t},
\eq
we arrive at the adjoint master equation for the operator $\hat O_t$.\\

The integral in Eq.~\eqref{001Eq.H.0.5.bis} depends on the choice of the kernel $\mathcal O(\lambda,\mu)$. On the other hand, we want an equation that can be directly applied to a generic operator $\hat O$ without having first to determine its kernel. This means that we want to rewrite Eq.~\eqref{001Eq.H.0.7.c} in the following time-dependent form
\bqali\label{adjoint.master}
\dfrac{d\hat \chi_t}{dt}&=\mathbb L_t\left[\hat \chi_t\right]=\dfrac{i}{\hbar}\left[\hat H_{\text{\tiny eff}}(t),\hat \chi_t\right]+\\
&+	\sum_{a,b=1}^2K_{ab}(t)\left[	\hat L_a	\hat \chi_t\hat L_b^{\dag}-\dfrac{1}{2}\left\{	\hat L_a\hat L_b^{\dag},\hat \chi_t	\right\}	\right],
\eqali
where the effective Hamiltonian $\hat H_{\text{\tiny eff}}(t)$, the hermitian Kossakowski matrix $\bm K(t)$ and the Lindblad operators $\hat L_a$ should not depend on the parameters $\lambda$ and $\mu$. Then, the linearity of Eq.~\eqref{001Eq.H.0.5.bis}
will allow to extend Eq.~\eqref{adjoint.master} to any operator $\hat O_t$. To achieve this, the explicit dependence from the parameters $\lambda$ and $\mu$, contained in the coefficients $\alpha_i$ and $\phi(t)$, must disappear in Eq.~\eqref{001Eq.H.0.7.c}. This can be done in the following way. Let us consider the commutation relations among $\hat x$, $\hat p$ and $\hat \chi_t$:
\bq\label{001Eq.H.0.6.bis}
\bigl[\hat \chi_t,\ \hat x	\bigr]=\hbar\alpha_2(t)\hat \chi_t\qquad\text{and}\qquad
\bigl[\hat \chi_t,\ \hat p	\bigr]=-\hbar \alpha_1(t)\hat \chi_t.
\eq
Given Eqs.~\eqref{001Eq.H.0.7.bis}, we can express $\lambda \hat \chi_t$ and $\mu \hat \chi_t$ as a linear combination of the above commutators. Then, by using this result, we can easily rewrite Eq.~\eqref{001Eq.H.0.7.c} in the form given by Eq.~\eqref{adjoint.master}, where
\bq\label{001Eq.G.3.29}
\hat H_{\text{\tiny eff}}(t)=\dfrac{\hat p^2}{2M}+\dfrac{\Gamma^{\text{\tiny A}}(t)}{2}\left(	\hat x\hat p+\hat p\hat x\right)+\dfrac{1}{2}M\Delta^{\text{\tiny A}}(t)\hat x^2,
\eq
the Lindblad operators are 
$\hat L_1=\hat x$ and $\hat L_2=\hat p$. The time dependent function $\Gamma^\text{\tiny A}(t)$, $\Delta^\text{\tiny A}(t)$ and the elements of the Kossakowski matrix $ K_{a,b}(t)$ are reported in Appendix \ref{app:B}.
An important note: one of the elements of the Kossakowski matrix vanishes $K_{22}(t)=0$. This means that the term corresponding to $[\hat p,[\hat p,\hat \rho_\text{\tiny S}]]$ is absent, as for the Caldeira-Leggett master equation \cite{Caldeira:1983aa}. In the latter case this implies the non complete positivity of the dynamics. In the case under study, complete positivity is instead automatically satisfied, as it is explicitly shown in Sec.~\ref{sec.cp}. This result is in agreement with previous results \cite{Haake:1985aa,Hu:1992aa,Ford:2001aa,Ferialdi:2016aa}.\\

Eq.~\eqref{adjoint.master} is linear in $\hat \chi_t$ and does not depend on $\lambda$ and $\mu$. Therefore because of Eq.~\eqref{001Eq.H.0.5.bis}, it holds for any operator $\hat O_t$: $\tfrac{d}{dt}\hat O_t=\mathbb L_t[\hat O_t]$, as in Eq.~\eqref{adjoint.master}.
This is the adjoint master equation and $\mathbb L_t$ is the generator of the dynamics. The correspondent adjoint dynamical map is given by
\bq\label{dynamical.map.operators}
{\bm \Phi_t}[\ \cdot\ ]=\mathcal T \exp\left(	\int_0^tds\, \mathbb L_s\right)[\ \cdot \ ].
\eq
The result here obtained is very general and depends only on the form of the total Hamiltonian $\hat H_\text{\tiny T}$ defining the QBM model together with the separability of the initial total state (Eq.~\eqref{001Eq.M.2}), but does not depend on the particular initial state of the system $S$. We now show that we recover the master equation \eqref{HPZmaster} for the states. \\

\section{The Master Equation for the statistical operator}
\label{subsec:HPZ}

We now derive the master equation for the density matrix, starting from the adjoint master equation. For a time independent adjoint master equation, switching to the master equation for the states is straightforward: the adjoint dynamical map ${\bm \Phi_t}$ is $\exp\left(t\mathbb L\right)$, where the generator $\mathbb L$ is time independent. Therefore the map ${\bm \Phi_t}$ and its generator $\mathbb L$ commute. Then the generator of the dynamics for the states  is equal to the adjoint of the generator of the dynamics for the operators. 
In the time dependent case here considered, instead, the procedure is more delicate. Consider the dynamical map ${\bm \Phi_t^*}$ for the states:
\bq\label{001Eq.HPZ.1.3}
{\bm \Phi_t^*}:\ 	\hat \rho_\text{\tiny S}(0)\mapsto\hat \rho_\text{\tiny S}(t),
\eq
which is the adjoint map of ${\bm \Phi_t}$ defined in Eq.~\eqref{dynamical.map.operators}. The adjointness, denoted here by the $*$-symbol, has to be understood in the following sense:
\bq
\braket{\hat \chi_t}=\TR{S}{{\bm \Phi_t}\left[\hat \chi(0)\right]\hat \rho_\text{\tiny S}(0)}=\TR{S}{\hat \chi(0){\bm \Phi_t^*}\left[\hat \rho_\text{\tiny S}(0)\right]},
\eq
where $\braket{\ \cdot\ }=\TR{S}{\ \cdot\ \hat \rho_\text{\tiny S}(0)}$.
Let us consider the time derivative of $\braket{\hat \chi_t}$ and let us express it as follows:
\bq\label{001Eq.HPZ.1.1}
\dfrac{d}{dt}\braket{\hat \chi_t}=\TR{S}{{\bm \Lambda_t}\left[\hat \chi(0)\right]\hat \rho_\text{\tiny S}(0)}=\TR{S}{\hat \chi(0){\bm \Lambda_t^*}\left[\hat \rho_\text{\tiny S}(0)\right]}.
\eq
The above equation defines the two maps ${\bm \Lambda_t}$ and ${\bm \Lambda_t^*}$. According to Eq.~\eqref{adjoint.master},
\bq\label{001Eq.HPZ.1.2}
{\bm \Lambda_t}\left[\hat \chi(0)\right]=\mathbb L_t\left[\hat \chi_t\right]=\mathbb L_t\circ{\bm \Phi_t}\left[\hat \chi(0)\right].
\eq
On the other hand, according to standard practice \cite{Breuer:2002aa}, the map ${\bm \Lambda_t^*}$ in Eq.~\eqref{001Eq.HPZ.1.1} is defined as
\bq\label{001Eq.HPZ.1.12}
{\bm \Lambda^*_t}\left[\hat \rho_\text{\tiny S}(0)\right]=\tilde{\mathbb L}^*_t\left[\hat \rho_\text{\tiny S}(t)\right]=\tilde{\mathbb L}^*_t\circ{\bm \Phi^*_t}\left[\hat \rho_\text{\tiny S}(0)\right],
\eq
where the map $\tilde{\mathbb L}^*_t$ is the generator of the dynamics for the states. By adjointness we have
\bq
\label{001.Eq.Lambdamapexp}
\TR{S}{\mathbb L_t\circ{\bm \Phi_t}\left[\hat \chi(0)\right]\ \hat \rho_\text{\tiny S}(0)}=\TR{S}{ {\bm \Phi_t}\circ\tilde{\mathbb L}_t\left[\hat \chi(0)\right]\hat \rho_\text{\tiny S}(0)}.
\eq
Then, by comparison we have to construct the map $\tilde{\mathbb L}_t$ as follows:
\bq\label{001Eq.HPZ.1.4}
\tilde{\mathbb L}_t={\bm \Phi_t}^{-1}\circ\mathbb L_t\circ{\bm \Phi_t}.
\eq
In terms of this latter expression, Eq.~\eqref{001Eq.HPZ.1.2} becomes
\bq\label{generator.strange}
{\bm \Lambda_t}\left[\hat \chi(0)\right]={\bm \Phi_t}\circ\tilde{\mathbb L}_t\left[\hat \chi(0)\right].
\eq

For a time dependent generator, in order to construct the master equation for the states we need to derive explicitly the form of $\tilde{\mathbb L}_t$. This is derived in Appendix \ref{app:C} and the final result is:
\bqali\label{001.Eq.Master}
\tilde{\mathbb L}_t\left[\hat \chi(0)\right]&=\dfrac{i}{\hbar}\left[\hat{\tilde H}_{\text{eff}}(t),\hat \chi(0)\right]+\\
&+	\sum_{a,b=1}^2\tilde K_{ab}(t)\left[	\hat L_a	\hat \chi(0)\hat L_b^{\dag}-\dfrac{1}{2}\left\{	\hat L_a\hat L_b^{\dag},\hat \chi(0)	\right\}	\right],
\eqali
where $\hat L_\alpha$ is defined after Eq.~\eqref{001Eq.G.3.29},  
\bq\label{001Eq.HPZ.1.14}
\hat{\tilde H}_{\text{eff}}(t)=\dfrac{\hat p^2}{2M}-\dfrac{\Gamma^{\text{\tiny A}}(t)}{2}\left(	\hat x\hat p+\hat p\hat x\right)+\dfrac{1}{2}M\Delta^{\text{\tiny A}}(t)\hat x^2,
\eq
and the elements of $\tilde K_{ab}(t)$ are reported in the Appendix.
Now, in order to obtain the time derivative of the operator $\hat O_t$ at time $t$, we act with $\tilde{\mathbb L}_t$  on the operator $\hat O(0)$ at time $t=0$ and then with the adjoint dynamical map ${\bm \Phi_t}$, as described in Eq.~\eqref{generator.strange}. The latter equation, according to the definition of the map ${\bm \Lambda_t}$ in Eq.~\eqref{001Eq.HPZ.1.2}, gives $\tfrac{d}{dt}\hat O_t$. From this we can compute the master equation for the states. This can be simply done by using the cyclic property of the trace $\TR{S}{\ \cdot \ }$ applied on the expression in Eq.~\eqref{001.Eq.Master}. Then, we have:
\bqali
\dfrac{d}{dt}\braket{\hat \chi_t}&=\TR{S}{\tilde{\mathbb L}_t\left[\hat \chi(0)\right]{\bm \Phi}_t^*\left[\hat \rho_\text{\tiny S}(0)\right]},\\&=\TR{S}{\hat \chi(0)\dfrac{d}{dt}\hat \rho_\text{\tiny S}(t)},
\eqali
which yields the master equation for the states of the system $S$:
\bqali\label{001Eq.HPZ.1.13}
\dfrac{d\hat \rho_\text{\tiny S}(t)}{dt}&=-\dfrac{i}{\hbar}\left[\hat{\tilde H}_{\text{eff}}(t),\hat \rho_\text{\tiny S}(t)\right]+\\
&+	\sum_{a,b=1}^2\tilde K_{ab}(t)\left[	\hat L_b^{\dag}	\hat \rho_\text{\tiny S}(t)\hat L_a-\dfrac{1}{2}\left\{	\hat L_a\hat L_b^{\dag},\hat \rho_\text{\tiny S}(t)	\right\}	\right],
\eqali
This is the desired result, which naturally coincides with the QBM master equation \eqref{HPZmaster}. The explicit form of the terms in Eq.~\eqref{001Eq.HPZ.1.13} can be obtained starting from the spectral density $J(\omega)$ defined in Eq.~\eqref{001Eq.SE.10}.

\section{Complete Positivity}
\label{sec.cp}

We now discuss the complete positivity of the dynamical map ${\bm \Phi_t}$ generated by the generator $\mathbb L_t$ defined in Eq.~\eqref{adjoint.master}. The action of this dynamical map on the generic operator $\hat O$ of the system $S$ is
\bq\label{001Eq.G.4.34}
{\bm \Phi_t}[\hat O]=\hat O_t=\TR{B}{\hat \rho_\text{\tiny B}\left(\hat{\mathcal U}^{\dag}_t(\hat O\otimes {\bf\hat1_B})\hat{\mathcal U}_t\right)},
\eq
which is the combination of two completely positive maps: the unitary evolution provided by the total Hamiltonian of system plus bath, and the trace over the bath. Therefore, by construction the dynamical map is completely positive. However, two observations are relevant here. First, it is instructive to verify explicitly the complete positivity of the dynamics. Second, in a situation where approximations are needed in order to compute explicitly the coefficients of the (adjoint) master equation, the verification of the complete positivity of the dynamics becomes a fundamental point of interest.\\

When the generator $\mathbb L$ of the dynamics is not time dependent, the sufficient and necessary condition for the complete positivity of the dynamical map is the positivity of the Kossakowski matrix \cite{Gorini:1976aa,Benatti:2009aa}. For a time dependent generator $\mathbb L_t$, instead, a positive Kossakowski matrix is only a sufficient condition for complete positivity. An example is precisely the QBM model under study, whose Kossakowski matrix is not positive for all times, nevertheless the dynamics is completely positive. For a time dependent generator, a necessary and sufficient condition instead is given by the following theorem \cite{Demoen:1977aa,Heinosaari:2010aa}, under the assumption of a Gaussian channel.\\

Suppose that the action of a gaussian dynamical map ${\bm \Phi_t}$ on the characteristic operator $\hat \chi$ of the system is defined as follows
\bq\label{001Eq.G.4.39}
{\bm \Phi_t}:		\exp\Bigl(i\braket{\xi|R}\Bigr)\mapsto \exp\left(i\braket{\xi|{\bm X}_t|R}\right)\exp\left(-\tfrac{1}{2}\braket{\xi|{\bm Y}_t|\xi}\right),
\eq
where ${\bm X}_t$ and ${\bm Y}_t$ are $2\times 2$ matrices describing the evolution of the characteristic operator 
\bq\label{001Eq.G.4.41}
\begin{matrix}
{\bm X}_t=
\begin{pmatrix}
G_1(t)		&G_2(t)/M\\
M\dot G_1(t)	&\dot G_2(t)
\end{pmatrix}
,&
{\bm Y}_t=
\begin{pmatrix}
-2\phi_1(t)		&-\phi_3(t)\\
-\phi_3(t)	&-2\phi_2(t)
\end{pmatrix},
\end{matrix}
\eq
 and $\bra{\xi}=(\lambda,\mu)$ and $\bra{R}=(\hat x,\hat p)$. In terms of ${\bm X}_t$, ${\bm Y}_t$ and of the symplectic matrix ${\bm \Omega}=\bigl(\begin{smallmatrix}0&&1\\\ -1&&0\end{smallmatrix} \bigr)$, we can define the following matrix ${\bm \Psi}_t$:
\bq\label{001Eq.G.4.42}
{\bm\Psi}_t={\bm Y}_t+\dfrac{i\hbar}{2}{\bm \Omega}-\dfrac{i\hbar}{2}{\bm X}_t{\bm \Omega}{\bm X}_t^{\tran}.
\eq
The necessary and sufficient condition for the dynamical map ${\bm \Phi_t}$ to be completely positive (CP) is the positivity of ${\bm \Psi}_t$ for all positive times.
Since the matrix ${\bm \Psi}_t$ is a $2\times2$ matrix, the request of its positivity reduces to the request of positivity of its trace and determinant:
\begin{subequations}\label{001Eq.G.4.44}
\begin{align}
\label{001Eq.G.4.44a}
\Tr{{\bm \Psi}_t}&=-2\Bigl(\phi_1(t)+\phi_2(t)\Bigr),\\
\label{001Eq.G.4.44b}
\Det{{\bm \Psi}_t}&=4\phi_1(t)\phi_2(t)-\phi_3^2(t)-\dfrac{1}{4}\Bigr(	\hbar-F(t)	\Bigl)^2.
\end{align}
\end{subequations}
The condition of positivity of the trace, Eq.~\eqref{001Eq.G.4.44a}, is easily verified for all physical spectral densities: the spectral density is positive by definition, see Eq.~\eqref{001Eq.SE.10}, and this implies the negativity of $\phi_1(t)$ and $\phi_2(t)$, see Eqs.~\eqref{001Eq.G.3.20}, for all positive values of the temperature. On the other hand, the second condition, Eq.~\eqref{001Eq.G.4.44b}, cannot be easily verified in general. Once a specific spectral density $J(\omega)$ is chosen, one can check explicitly whether $ \Det{{\bm \Psi}_t}\geq0$. For example, the spectral density $J(\omega)\propto\omega$, originally chosen in \cite{Caldeira:1983aa} to describe the quantum brownian motion, does not satisfy the above condition also in the case of no external potentials, and in fact it is well known that the Caldeira-Leggett master equation is not CP.\\

As already remarked, the QBM model automatically guarantees complete positivity. However, in practical cases one is not able to compute explicitly the time dependent coefficients of the Kossakowski matrix. Approximations are needed, in which case complete positivity is not automatically guaranteed anymore. This can be checked in a relatively easy way by assessing the positivity of $\Det{{\bm \Psi}_t}$.

\section{Time evolution of relevant quantities}
\label{general_sol}

The original QBM master equation~\eqref{HPZmaster} is expressed in terms of functions (forming the Kossakowski matrix), whose explicit expression is not easy to derive, even if one considers the solution given in \cite{Ford:2001aa}. They are solutions of complicated differential equations, difficult to solve except for very simple situations. More important, expectation values are not easy to compute: one has to determine the state of the system at time $t$, which is in general a formidable problem also in a particularly simple situation. In our derivation, instead, the use of the adjoint master equation provides an much easier tool for the computation of expectation values. The evolution is expressed in the Heisenberg picture, therefore it does not depend on the state of the system $S$ but only on the properties of the adjoint evolution ${\bm \Phi_t}$.

For example, by plugging the expression of $\hat x^2(t)$ (obtained from Eq.~\eqref{001Eq.H.0.6b}) in Eq.~\eqref{adjoint.master} we obtain an equation for the expectation value $\braket{\hat x^2_t}$:
\bq\label{differentialeq}
\begin{aligned}
&\dfrac{d}{dt}\braket{\hat x^2_t}=2\dot G_1(t)\dot G_2(t)\braket{\hat x^2}+2\dot
G_1(t)\dot G_2(t)\braket{\hat p^2}/M^2+\\
&+(G_1(t)\dot G_2(t)+\dot G_1(t)G_2(t))\braket{\{\hat x,\hat p\}}/M-2\dot\phi_1(t),
\end{aligned}
\eq
which can be solved directly
without having to solve a 
more complicated system of differential equations, as  it is 
necessary when the solution is in the Schr\"odinder picture \cite{Breuer:2002aa}, as well as for the case of the Wigner function approach \cite{Halliwell:1996aa,Halliwell:2007aa,Ford:2001aa}.
Once the interaction with the bath, i.e.~the spectral density
function, is specified, $G_1(t)$ and $G_2(t)$ can be determined as
described before, and this fully determines the time evolution of
$\braket{\hat x^2_t}$ in terms of the initial expectation values. In a similar way one can compute all the other expectation values as a function of time.

 To show this, we provide the explicit general solution of some physical quantities of interest for a specific spectral density. We consider: the diffusion function $\Lambda^\text{\tiny dif}(t)=\braket{\hat x^2_t}-\braket{\hat x_t}^2$, the energy of the system $E(t)=\braket{\hat p^2_t}/2M+\tfrac{1}{2}M\omega_\text{\tiny S}^2\braket{\hat x_t^2}$ and the decoherence function $\Gamma_{dec}(t)$. The latter is defined as follows.
We consider a particle which, at time $t=0$, is described by a state $\Ket{\psi(t=0)}=\mathcal{N}[\Ket \alpha+\Ket \beta]$, where $\Ket \alpha$ and $\Ket \beta$ are two equally spread out, with spread equal to $\sigma_0$, gaussian wave packets, centered respectively in $x_{\alpha}=\braket{\alpha|\hat x|\alpha}$ and $x_{\beta}=\braket{\beta|\hat x|\beta}$ and $\mathcal N$ is the normalization constant. The probability density in position $x$ at time $t$ is \cite{Breuer:2002aa}:
\bqali\label{001Eq.G.6.interfpattern}
\mathcal P(x,t)&=\mathcal N^2\left\{	\rho_{\alpha\alpha}(x,t)+\rho_{\beta\beta}(x,t)+\right.\\&\left.+2\sqrt{\rho_{\alpha\alpha}(x,t)\rho_{\beta\beta}(x,t)}\exp\left[\Gamma^\text{\tiny dec}(t)	\right]\cos\bigl[\varphi(x,t)	\bigr]	\right\},
\eqali
where $\rho_{\alpha\beta}(x,t)=\braket{x|\TR{B}{\mathcal U_t(\Ket{\alpha}\Bra{\beta})\mathcal U_t^{\dag}}|x}$: there is a modulation given by the phase $\varphi(x,t)$ and a reduction of the interference contrast determined by the decoherence function $\Gamma^\text{\tiny dec}(t)<0$. The decoherence function takes the following form:
\bqali\label{001Eq.G.6.decoherence}
\Gamma^\text{\tiny dec}(t)&=-\dfrac{4\sigma_0^4\Delta_p^2+\hbar^2\Delta_x^2}{\hbar^2}\cdot\\&\cdot\dfrac{M^2\phi_1(t)}{8M^2\sigma_0^2\phi_1(t)-\hbar^2G_2^2(t)-4M^2\sigma_0^4G_1^2(t)},
\eqali
where $\Delta_x$ and $\Delta_p$ are the distances between the two guassians in position and momentum, and the function $\phi_1(t)$ is defined in Eq.~\eqref{eq:defphi1}. The explicit expressions for $\Lambda^\text{\tiny dif}(t)$ and $E(t)$ is given in Appendix \ref{app:d}.

As a concrete example, we consider the case of the Drude-Lorentz
spectral density
\bq
J(\omega)=\frac{2}{\pi}M \gamma
\Omega^2\frac{\omega}{(\omega^2+\Omega^2)},
\eq
 which is commonly used
for example in light-harvesting systems
\cite{Wendling:2000aa,Fassioli:2012aa}, where $\Omega$ is the characteristic frequency of the bath. The corresponding dissipation and noise kernels, defined in Eq.~\eqref{001Eq.SE.14a} and Eq.~\eqref{001Eq.SE.14b} respectively, are:
\begin{subequations}
\bq
D(t)=2M \gamma\hbar\Omega^2 e^{-\Omega |t|}\text{sign}(t),
\eq
\bqali
&D_1(t)=\dfrac{2M \gamma\hbar\Omega^2}{\pi}\left[\Phi_L\left(e^{\tfrac{-2\pi |t|}{\beta\hbar}},1,-\dfrac{\beta\hbar\Omega}{2\pi}\right)	+\right.\\&\left.+\Phi_L\left(e^{\tfrac{-2\pi |t|}{\beta\hbar}},1,\dfrac{\beta\hbar\Omega}{2\pi}\right)\right]+2M\gamma\hbar\Omega^2 e^{-\Omega |t|}\cot\left(\dfrac{\beta\hbar\Omega}{2}\right),
\eqali
\end{subequations}
where the function $\Phi_L$ is the Hurwitz-Lerch transcendent function
$\Phi_L(z,s,a)=\sum_{n=0}^{+\infty}z^n(n+a)^{-s}$.
The two Green functions are:
\bq
G_2(t)=\sum_{i=1}^3 \dfrac{(\Omega+ C_i )e^{C_i t}}{D_i},
\eq
and $G_1(t)=\tfrac{d}{dt}G_2(t)$,
where $C_1$, $C_2$ and $C_3$ are the complex roots of the polynomial $y(s)=(y^2+\omega_\text{\tiny S}^2+2\gamma \Omega)(y+\Omega)-2\gamma\Omega^2$ and $D_i=\prod_{j=1, j\neq i}^{3}(C_i-C_j)$. In terms of these functions, we can compute the functions $\phi_i(t)$ with the help of Eqs.~\eqref{001Eq.G.3.20} as well as the three relevant quantities previously discussed, whose explicit expressions are displayed in Eq.~\eqref{001Eq.G.6.decoherence}, Eq.~\eqref{001Eq.G.6.diffus} and Eq.~\eqref{001Eq.G.6.energy}.\\

\begin{figure}[t!]
\centerline{\includegraphics[width=1\linewidth]{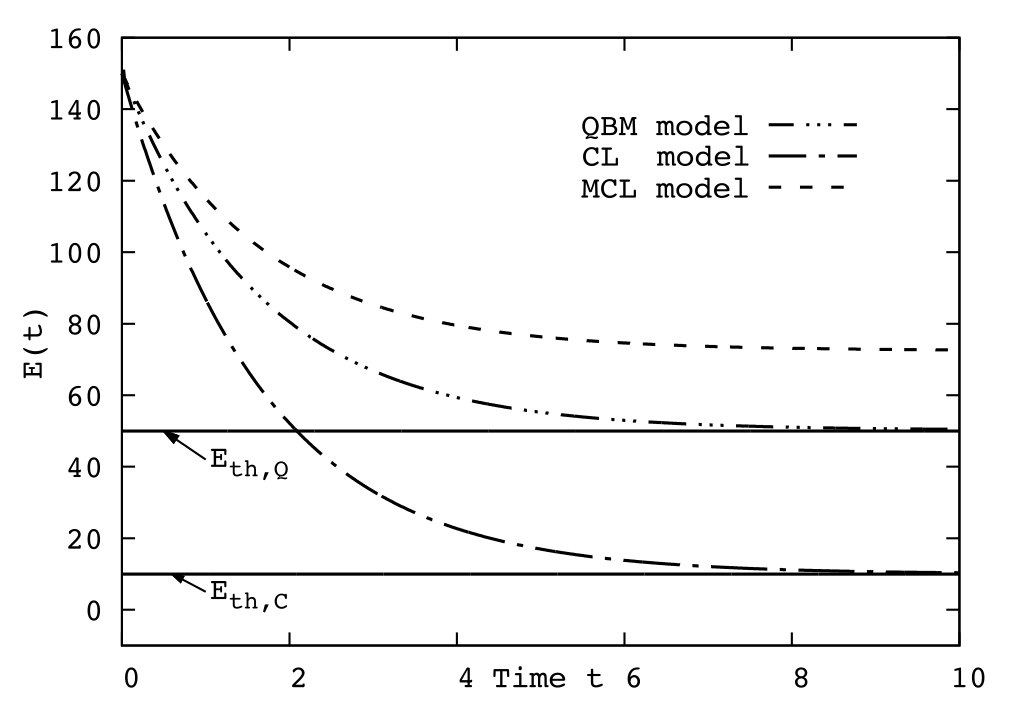}}
\centerline{\includegraphics[width=1\linewidth]{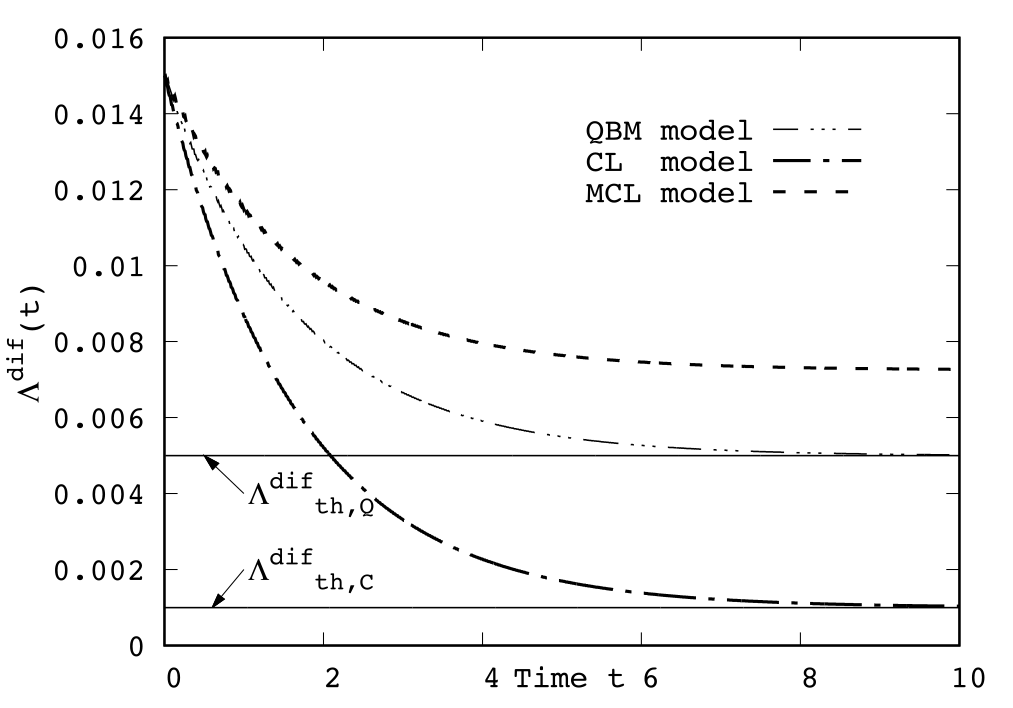}}
\caption{Time evolution of the energy $E(t)$ (top panel)
 and diffusion in space $\Lambda^\text{\tiny dif}(t)$ (bottom panel)
 for the first excited state of
the harmonic oscillator with frequency $\omega_\text{\tiny S}=100$ centred in the
origin ($\braket{\hat x}=0=\braket{\hat p}$) with parameters $M=1$, $\gamma=0.3$, $\Omega=2000$, $\beta=10^{-1}$ and $\hbar=1$. The plot shows the behavior of $E(t)$
and $\Lambda^\text{\tiny dif}(t)$ for the QBM model with the Drude-Lorentz spectral
density, for the Caldeira-Leggett model (CL) and for its modification (MCL).  $E_\text{\tiny th,Q}$, $E_\text{\tiny th,C}$,
$\Lambda^\text{\tiny dif}_\text{\tiny th,Q}$ and $\Lambda^\text{\tiny dif}_\text{\tiny th,C}$ (see main text) are also plotted.}
\label{high_energy}
\end{figure}

\begin{figure}[t!]
\centerline{\includegraphics[width=1\linewidth]{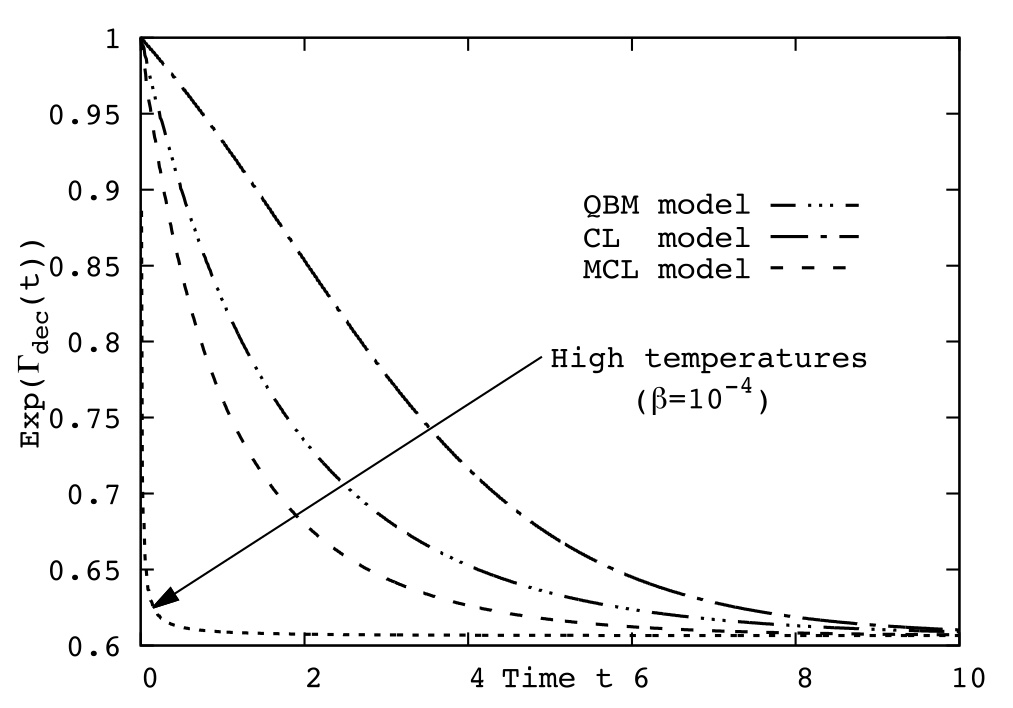}}
\caption{Decoherence function $\exp(\Gamma^\text{\tiny dec}(t))$ with parameters
$M=1$, $\gamma=0.3$, $\Omega=200$, $\omega_\text{\tiny S}=100$, $\hbar=1$,
$\sigma_0=\sqrt{\hbar/(2M\omega_\text{\tiny S})}$, $\Delta_x=2\sigma_0$ and
$\Delta_p=0$.  The plot shows the behavior of the decoherence function
$\exp(\Gamma(t))$ for the Drude-Lorentz spectral density (QBM), for
the Caldeira-Leggett model (CL) and for its modification (MCL) at two different temperatures: $\beta=10^{-1}$ and
$\beta=10^{-4}$. For $\beta=10^{-4}$ the differences between the three models are minimal and the three curves coincide with the dotted line.}
\label{fig.decoh}
\end{figure}

Fig.~\ref{high_energy} and \ref{fig.decoh} show the evolution of the diffusion function $\Lambda^\text{\tiny dif}(t)$, of the energy $E(t)$ and of the decoherence function $\Gamma^\text{\tiny dec}(t)$, and we
compare their time evolution according to the QBM model as described here above (QBM),
with that of the Caldeira-Leggett (CL) master equation in
\eqref{001Eq.M.1.3}.  We also consider the evolution given by the
Modification of the Caldeira-Leggett (MCL) master equation,
which is obtained from Eq.~\eqref{001Eq.M.1.3} by adding the term
$-\tfrac{\gamma \beta}{8M}[\hat p,[\hat p,\hat \rho_\text{\tiny S}(t)]]$ to
guarantee the complete positivity of the dynamics
\cite{Ambegaokar:1991aa,Diosi:1993aa,Breuer:2002aa}.  As for the initial state,
in Fig.~\ref{high_energy} we considered the first excited state of
the harmonic oscillator with frequency $\omega_\text{\tiny S}$ centred in the
origin: $\braket{\hat x}=0=\braket{\hat p}$. 

The asymptotic value of $E(t)$ is given by the equilibrium energy of
the thermal state $\hat \rho_{th}\propto \exp(-\beta H_\text{\tiny S})$:
\bq\label{Eq.energy.th} E_\text{\tiny th,Q}=\dfrac{\hbar
\omega_\text{\tiny S}}{2}+\dfrac{\hbar \omega_\text{\tiny S}}{e^{\beta \hbar \omega_\text{\tiny S}}-1}, \eq
which the high temperature limit coincides with the classical value 
$E_\text{\tiny th,C}=1/\beta$. 
For high temperatures the difference between the
two thermal energies, $E_\text{\tiny th,Q}$ and $E_\text{\tiny th,C}$, is negligible; in
this case the three dynamics lead to the same asymptotic value.
This is expected since both CL and MCL are derived in the high
temperature limit, and our result is exact.  However at low temperatures, as
Fig.~\ref{high_energy} shows, the difference between the quantum and
classical case becomes important and shows the quantum properties of
the system $S$: the zero-point energy $\hbar\omega_\text{\tiny S}/2$ is the minimal
allowed energy.  The CL dynamics, at low temperatures, fails to
capture this feature since its asymptotic value is lower.  The MCL
dynamics leads to an asymptotic energy which is different from both
the classical and the quantum value.  This is due to the correction to
the Caldeira-Leggett master equation.  As mentioned before, the latter
is needed to satisfy complete positivity, however it leads to
unphysical effects, e.g.~the system is overheated.  Only the QBM model
displays the correct quantum behavior.

A similar situation is found for the diffusion in position
$\Lambda^\text{\tiny dif}(t)$.  According to the well-known result of equilibrium
quantum statistical physics, its asymptotic value is given by
\cite{Feynman:1965aa,Caldeira:1983aa}:
\bq\label{Eq.diffus.th} \Lambda^\text{\tiny dif}_\text{\tiny th,Q}=\dfrac{\hbar}{2M
\omega_\text{\tiny S}}\coth\left(\dfrac{\beta\hbar\omega_\text{\tiny S}}{2} \right), \eq
which is the diffusion for an harmonic oscillator in the thermal state
$\hat \rho_{th}$.  In the high
temperature limit  Eq.~\eqref{Eq.diffus.th} gives the classical
asymptotic value
$
\Lambda^\text{\tiny dif}_\text{\tiny th,C}=1/M \beta \omega_\text{\tiny S}^2$.
Again, for high temperatures the difference between the classical and
quantum thermal diffusion can be neglected, and the three dynamics
give the same result.  For low temperatures the difference becomes
important.  The MCL asymptotic value differs both from the classical
and quantum equilibrium values.\\
Fig.~\ref{fig.decoh} shows how $\Gamma^\text{\tiny dec}(t)$ decays in time.
For high temperatures, $\exp(\Gamma^\text{\tiny dec}(t))$ reaches rapidly its
asymptotic value, i.e.~the decoherence time $\tau_\text{\tiny D}$ is very short.
In the low temperature case instead $\tau_\text{\tiny D}$ is higher.  Notice that
the asymptotic value in both cases is not zero but, in agreement with
the literature \cite{Breuer:2002aa}, it saturates at a finite value:
\bq
\Gamma^\text{\tiny dec}(\infty)=-\frac{1}{8}\frac{\Delta_x^2}{\sigma_0^2}.
\eq
  Again,
there are differences between the three dynamics.  In particular, with respect to the QBM result, the
CL dynamics overestimate the decoherence time $\tau_\text{\tiny D}$ whereas for the
MCL it is underestimated.

\subsection{Non-gaussian initial state}
\begin{figure}[b!]
\centerline{\includegraphics[width=1\linewidth]{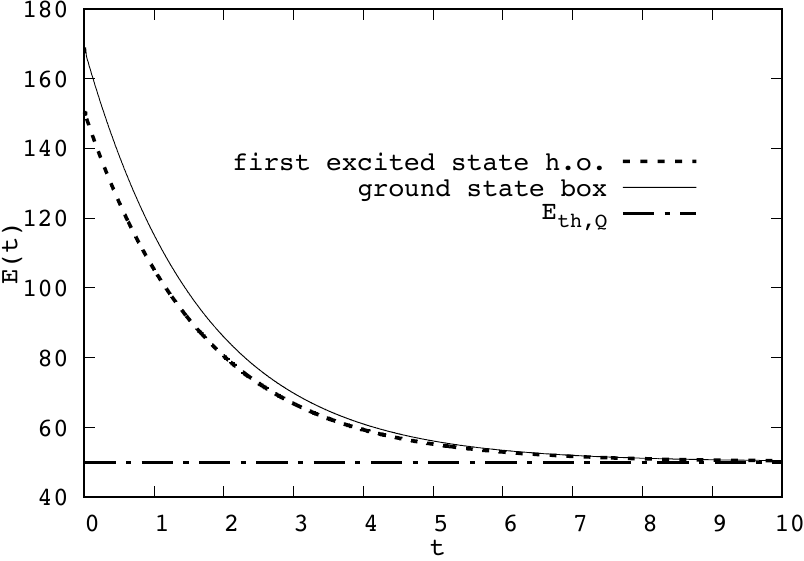}}
\centerline{\includegraphics[width=1\linewidth]{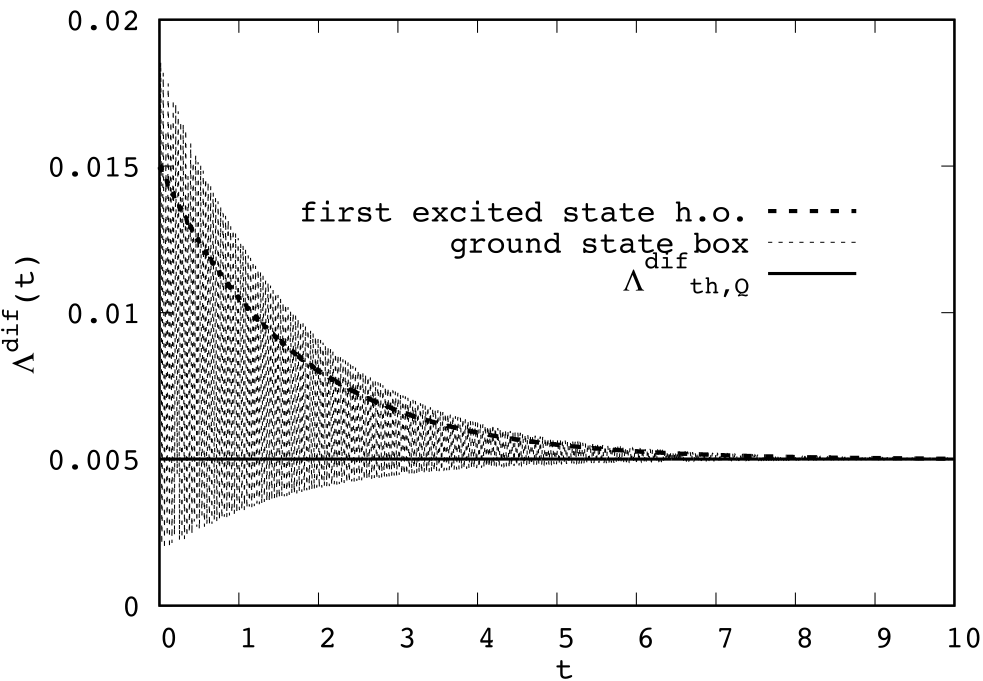}}
\caption{Comparison of the solutions of the QBM model for two different initial states: the first excited state of the harmonic oscillator at frequency $\omega_\text{\tiny S}$ (dashed line) and the ground state of the square potential displayed in Eq.~\eqref{state.box} (continuous line). The chosen parameters are: $M=1$, $\gamma=0.3$, $\Omega=2000$,
$\omega_\text{\tiny S}=100$, $\beta=10^{-1}$, $a=0.23$ and $\hbar=1$. Top panel: Evolution of the energy $E(t)$ for the two systems, compared with the equilibrium energy $E_\text{\tiny th,Q}$ for the quantum thermal state. Bottom panel: Evolution of the diffusion function $\Lambda^\text{\tiny diff}(t)$ for the two systems, compared with the equilibrium diffusion $\Lambda^\text{\tiny diff}_\text{\tiny th,Q}$ for the quantum thermal state.}
\label{fig:compare}
\end{figure}
The following example will make clear the advantage of the
present approach. Consider a system initially confined by the square potential 
$V(x)=0$ for $x\in[0,a]$ and $V(x)=+\infty$ otherwise, at rest in the ground state. The system later evolves subject to the harmonic potential. The initial state then is:
\bq\label{state.box}
\psi(x)=
\begin{cases}
\sqrt{2/a}\sin(\pi x/a)& x\in[0,a],\\
0& \text{otherwise}.
\end{cases}
\eq
The corresponding initial expectation values for the quadratic operators are:
 \bq\label{values.square}
 \begin{split}
 \braket{\hat x}=\tfrac a2,&\quad
 \braket{\hat p}=0,\\
 \braket{\hat x^2}=\tfrac16(2-\tfrac{3}{\pi^2})a^2,\quad
  \braket{\hat p^2}&=\tfrac{\pi^2\hbar^2}{a^2},\quad
  \braket{\{\hat x,\hat p\}}=0.
\end{split}
\eq
The time evolution of the diffusion function $\Lambda^\text{\tiny dif}(t)$ and energy $E(t)$ is easy to obtain, as one can see from Eq.~\eqref{001Eq.G.6.diffus} and Eq.~\eqref{001Eq.G.6.energy}.
In fact in our approach the only quantities that might change, when changing the state of the system, are the initial expectation values. The functional dependence of the physical quantities on the initial values instead does not change. Then, by plugging in Eq.~\eqref{001Eq.G.6.diffus} and Eq.~\eqref{001Eq.G.6.energy} the initial expectation values for the non-gaussian state (Eq.~\eqref{values.square}), one directly obtains the time evolution of $\Lambda^\text{\tiny dif}(t)$ and $E(t)$, which are plotted \footnote{The ground state of the square potential was preferred to its first excited since the initial values of energy and diffusion function are more compatible with those of the first excited state of the harmonic oscillator. However, a similar time dependence is shown when the initial state is taken equal to the first excited state of the square potential.} in Fig.~\ref{fig:compare}. While the time evolution of $E(t)$ is qualitatively the same as in the example previously considered, the diffusion function $\Lambda^\text{\tiny dif}(t)$ shows high frequency oscillations when the initial state is taken equal to Eq.~\eqref{state.box}. These oscillations arise from the choice of the initial state and are present also when the system is isolated.

With no bath, the diffusion function is equal to
\bqali\label{lambdafree}
\Lambda^\text{\tiny dif}(t)&=\cos^2\omega_\text{\tiny S}t\left(\braket{\hat x^2}-\braket{\hat x}^2\right)+\frac{\sin^2\omega_\text{\tiny S}t}{M^2\omega_\text{\tiny S}^2}\left(\braket{\hat p^2}-\braket{\hat p}^2\right)\\
&+\frac{2\cos\omega_\text{\tiny S}t\sin\omega_\text{\tiny S}t}{M \omega_\text{\tiny S}}\left(\frac{\braket{\{\hat x,\hat p\}}}{2}-\braket{\hat x}\braket{\hat p}\right).
\eqali 
By plugging into this expression the expectation values for the ground state of the square potential (see Eq.~\eqref{values.square})  we obtain the oscillatory behaviour, while for the eigenstates of the harmonic oscillator Eq.~\eqref{lambdafree} the diffusion of course is constant (and correspondingly when the bath is switched on, $\Lambda^\text{\tiny dif}(t)$ simply decays exponentially as plotted in Fig.~\ref{fig:compare}).

As we have shown, the evolution of the expectation values is easy to obtain by using our approach. Once the functional dependence of the physical quantities on the initial values is computed, we direct obtain their time dependence for different initial states simply by inserting the initial expectation values.
 On the other hand, when working in the Schr\"odinger picture, as typically done in the literature \cite{Hu:1992aa,Breuer:2002aa}, or with the Wigner formalism \cite{Halliwell:1996aa,Halliwell:2007aa,Ford:2001aa}, one has to find the explicit time evolution of the initial state, which changes depending on the initial state.

\section{Conclusions}

We described an alternative approach to  Quantum Brownian motion, based on the Heisenberg picture. The essential ingredients are three: \emph{i)} the full Hamiltonian \eqref{001Eq.M.1}, describing both the evolution of the system and of the bath, \emph{ii)} an uncorrelated initial state for the system and the bath \eqref{001Eq.M.2}, and \emph{iii)} the spectral density \eqref{001Eq.SE.10}, which has to  satisfy precise physical constraints \footnote{The most important physical constrain on the spectral density is dictated by the validity of the Fluctuation-Dissipation theorem \eqref{001Eq.SE.16} for the dissipative $D(t)$ and noise $D_1(t)$ kernels.}.

Starting from these ingredients, we derived explicitly the adjoint master equation \eqref{adjoint.master} for a generic operator of the system. 
Due to the specific structure of the characteristic operator,
from the adjoint master equation we obtained the more familiar master equation for the statistical operator \eqref{001Eq.HPZ.1.13}. In general, this procedure is not straightforward, however in this case it was possible to carry out the calculations analytically. As expected, the master equation we obtain is equivalent to previous results \cite{Haake:1985aa,Hu:1992aa}.

 A criterion for the complete positivity of the dynamics is given. This becomes important when approximations are needed to carry out calculations and then complete positivity is not guaranteed anymore.
 
The two approaches (Heisenberg and Schr\"odinger) are equivalent, however the explicit expression of the coefficients of the master equation, in the original framework of Eq.~\eqref{HPZmaster}, can be given only in the weak coupling regime \cite{Hu:1992aa}, whereas for the approach here presented it can be given for more general and physically relevant situations \cite{Diosi:2014aa}. A similar result was obtained in \cite{Ford:2001aa}, however there is an important difference with respect to our approach: differently from \cite{Ford:2001aa} we are not bound to computing the time evolution of the state of the system, which in general is a complicated task. The explicit dependence from the initial state appears only in the initial expectation values, and not in the dynamics. This simplifies the derivation of expectation values of physical quantities and, even more, it makes the latter possible also for non trivial states such as gaussian state. \\

\section{Acknowledgements}

The authors acknowledge financial support from the University of Trieste (FRA 2016) and INFN.

\appendix

\section{Explicit form of $\phi(t)$}
\label{app:A}
With reference to Eq.~\eqref{eq:17}, it is easy to see that for $\hat \rho_\text{\tiny B}$ as in Eq.~\eqref{eq:18} the trace over $\hat \chi_{\beta}(t)$ gives a real and positive function of time.
Using the definition of the spectral density in Eq.~\eqref{001Eq.SE.10} one immediately derives: $\phi(t)=\lambda^2\phi_1(t)+\mu^2\phi_2(t)+\lambda\mu\phi_3(t)$, where the explicit form of $\phi_i(t)$ is:
\begin{subequations}\label{001Eq.G.3.20}
\begin{align}
\phi_1(t)&=-\int_0^tdt'\int_0^tds\ \dfrac{D_1(t'-s)}{4M^2}G_2(t-t')G_2(t-s),\label{eq:defphi1}\\
\phi_2(t)&=-\dfrac{1}{4}\int_0^tdt'\int_0^tds\ D_1(t'-s)G_1(t-t')G_1(t-s),\\
\phi_3(t)&=-\dfrac{G_2(t)}{2M}\int_0^tds D_1(s)G_2(t-s),
\end{align}
\end{subequations}
with
\bq\label{001Eq.SE.14b}
D_1(t)=2\hbar\int_0^{+\infty}d\omega\
J(\omega)\coth(\beta\hbar\omega/2)\cos(\omega t),
\eq
denoting the noise kernel. $D_1(t)$ is related to the dissipative kernel $D(t)$ through the Fluctuation-Dissipation theorem \cite {Callen:1951aa,Callen:1952aa,Kubo:1966aa,Pottier:2001aa,Yan:2005aa}:
\bqali\label{001Eq.SE.16}
\int_{-\infty}^{+\infty}dt&\ \cos(\omega t)D_1(t)=\\
&\coth\left(\dfrac{\beta\hbar\omega}{2}\right)\int_{-\infty}^{+\infty}dt\ \sin(\omega t) D(t).
\eqali

\section{Explicit form of the adjoint master equation}
\label{app:B}

Starting from Eqs.~\eqref{001Eq.H.0.7.bis} for $\alpha_1(t)$ and $\alpha_2(t)$, linear combinations of these relations give the following relations:
\begin{subequations}\label{001Eq.G.3.22}
\begin{align}
\label{001Eq.H.0.7.ab}
\lambda \hat \chi_t&=-\dfrac{M\dot G_1(t)}{F(t)}\left[\hat \chi_t,\hat x\right]-\dfrac{\dot G_2(t)}{F(t)}\left[\hat \chi_t,\hat p\right],
\\
\label{001Eq.H.0.7.aa}
\mu \hat \chi_t&=\dfrac{G_1(t)}{F(t)}\left[\hat \chi_t,\hat x\right]+\dfrac{G_2(t)}{MF(t)}\left[\hat \chi_t,\hat p\right],
\end{align}
\end{subequations}
where we defined
\bq\label{001Eq.G.3.23}
F(t)=\hbar\left(G_1(t)\dot G_2(t)-\dot G_1(t)G_2(t)\right).
\eq
By combining the results in Eq.~\eqref{001Eq.H.0.7.c} and Eqs.~\eqref{001Eq.G.3.22}, one immediately can check that Eq.~\eqref{001Eq.H.0.7.c} takes the Lindblad time-dependent form described in Eq.~\eqref{adjoint.master}. In particular, the effective Hamiltonian $\hat H_{\text{\tiny eff}}$ is given by Eq.~\eqref{001Eq.G.3.29},
where
\begin{subequations}\label{001Eq.G.3.30}
\begin{align}
\Gamma^{\text{\tiny A}}(t)&= \dfrac{\hbar}{2}\dfrac{\left(G_1(t)\ddot G_2(t)-\ddot G_1(t)G_2(t)\right)}{F(t)},
\\
\Delta^{\text{\tiny A}}(t)&=\hbar\dfrac{\left(	\dot G_1(t)\ddot G_2(t)-\ddot G_1(t)\dot G_2(t)\right)}{F(t)},
\end{align}
\end{subequations}
and the elements of the Kossakowski matrix $ K_{a,b}(t)$ are:
\bq\label{001eq.HPZ.1.11} \begin{aligned} K_{11}(t)&=\dfrac{1}{\hbar
F(t)}\int_0^tds \ D_1(s)\left( G_1(t)\dot G_2(t-s)+\right.\\
&\left.-\dot G_1(t)G_2(t-s)\right),\\ K_{12}(t)&=\dfrac{1}{2M\hbar
F(t)}\int_0^tds\ D_1(s)\Bigl( G_1(t-s)G_2(t)+\Bigr.\\
&\Bigl.-G_1(t)G_2(t-s)\Bigr)-i\dfrac{\Gamma^{\text{\tiny A}}(t)}{\hbar}, \end{aligned}
\eq
and $ K_{22}(t)=0$.

\section{Derivation of the master equation for the states}
\label{app:C}

To construct $\tilde{\mathbb L}_t$, we start from the derivative with respect to the parameters $\lambda$ and $\mu$ of the characteristic operator $\hat \chi_t$, see Eq.~\eqref{001Eq.H.0.7.1} of the main text:
\bq
\label{001Eq.HPZ.1.5a}
\dfrac{\partial}{\partial \lambda}\hat \chi_t=i G_1(t)\hat x \hat \chi_t+i\dfrac{G_2(t)}{M}\hat p\hat \chi_t+A(t)\hat \chi_t,
\eq
and 
\bq
\label{001Eq.HPZ.1.5b}
\dfrac{\partial}{\partial \mu}\hat \chi_t=i M \dot G_1(t)\hat x\hat \chi_t+i \dot G_2\hat p\hat \chi_t+B(t)\hat \chi_t,
\eq
where:
\bqali\label{001Eq.HPZ.1.6}
A(t)&=\left(\dfrac{i}{2} F(t)+\phi_3(t)\right)\mu+2\phi_1(t)\lambda,
\\
B(t)&=\left(-\dfrac{i}{2}F(t)+\phi_3(t)\right)\lambda+2\phi_2(t)\mu,
\eqali
where $F(t)$ is defined in Eq.~\eqref{001Eq.G.3.23}.
By linearly combining Eqs.~\eqref{001Eq.HPZ.1.5a} and \eqref{001Eq.HPZ.1.5b} we arrive at the following expressions:
\bqali\label{001Eq.HPZ.1.7}
\hat x\hat \chi_t&=\dfrac{i\hbar}{M F(t)}G_2(t)\dfrac{\partial}{\partial \mu}\hat \chi_t-\dfrac{i \hbar}{F(t)}\dot G_2(t)\dfrac{\partial}{\partial \lambda}\hat \chi_t+\\
&+\dfrac{i\hbar}{F(t)}\left[\dot G_2
(t)A(t)-\dfrac{G_2(t)}{M}B(t)	\right]\hat \chi_t,\\
\hat p\hat \chi_t&=-\dfrac{i\hbar}{F(t)}G_1(t)\dfrac{\partial }{\partial \mu}\hat \chi_t+\dfrac{i \hbar}{F(t)}M\dot G_1(t)\dfrac{\partial}{\partial \lambda}\hat \chi_t+\\
&-\dfrac{i\hbar}{F(t)}\left[	M\dot G_1(t)A(t)-G_1(t)B(t)\right]\hat \chi_t,
\eqali
which we use to replace the terms proportional to $\hat x$ and $\hat p$ in Eq.~\eqref{001Eq.H.0.7.c} of the main text; the right-hand side gives $\mathbb L_t[\hat \chi_t]=\mathbb L_t\circ{\bm \Phi_t}[\hat \chi(0)]$. By multiplying it from the left with ${\bm \Phi}_t^{-1}$ we obtain
\bqali\label{001Eq.HPZ.1.8}
{\bm \Phi}_t^{-1}\circ&\, \mathbb L_t\circ{\bm \Phi_t}\left[\hat \chi(0)\right]=\tilde{\mathbb L}_t\left[\hat \chi(0)\right]=\\
&={\bm \Phi}_t^{-1}\left[	
M \Delta^{\text{\tiny A}}(t)\mu \left(A(t)-\dfrac{\partial}{\partial\lambda}\right)+\right.\\
&\left.+\left(\dfrac{\lambda}{M}+2\mu \Gamma^{\text{\tiny A}}(t)\right)\left(\dfrac{\partial}{\partial\mu}-B(t)\right)\right.+\\
&\left.+\dfrac{i\hbar}{2}\bigl[	\dot\alpha_1(t)\alpha_2(t) - \alpha_1(t)\dot\alpha_2(t)\bigr]+\dot\phi(t)\right]{\bm \Phi_t}\left[\hat \chi(0)\right],
\eqali
where $\Delta^{\text{\tiny A}}(t)$ and $\Gamma^{\text{\tiny A}}(t)$ are defined in Eqs.~\eqref{001Eq.G.3.30} of the main text. Now, the expression within the square brackets contains no operator, therefore the action of the inverse map ${\bm \Phi}_t^{-1}$ and of the direct map ${\bm \Phi}_t$ cancel each other. Moreover, because of Eqs.~\eqref{001Eq.HPZ.1.5a} and \eqref{001Eq.HPZ.1.5b} we have
\bqali\label{001Eq.HPZ.1.9}
\dfrac{\partial}{\partial \lambda}\hat \chi(0)&=i\left(\hat x+\dfrac{\hbar}{2}\mu\right)\hat \chi(0),
\\
\dfrac{\partial}{\partial \mu}\hat \chi(0)&=i\left(\hat p-\dfrac{\hbar}{2}\lambda\right)\hat \chi(0),
\eqali
and therefore Eq.~\eqref{001Eq.HPZ.1.8} becomes
\begin{multline}\label{001Eq.HPZ.1.8second}
\tilde{\mathbb L}_t\left[\hat \chi(0)\right]=\left[	
M \Delta^{\text{\tiny A}}(t)\mu \left(A(t)-i\left(	\hat x+\dfrac{\hbar}{2}\mu\right)\right)+\right.\\
\left.+\left(\dfrac{\lambda}{M}+2\mu \Gamma^{\text{\tiny A}}(t)\right)\left(i\left(	\hat p-\dfrac{\hbar}{2}\lambda\right)-B(t)\right)\right.+\\
\left.+\dfrac{i\hbar}{2}\bigl[	\dot\alpha_1(t)\alpha_2(t) - \alpha_1(t)\dot\alpha_2(t)\bigr]+\dot\phi(t)\right]\hat \chi(0).
\end{multline}
Now we want to rewrite the above relation without any explicit dependence on $\lambda$ and $\mu$. In order to do so, we use the same procedure used in passing from Eq.~\eqref{001Eq.H.0.7.c} to Eq.~\eqref{adjoint.master} of the main text. According to Eq.~\eqref{001Eq.H.0.6.bis}:
\bq\label{001Eq.HPZ.1.10}
\bigl[\hat \chi(0),\ \hat x	\bigr]=\hbar\mu\hat \chi(0)\qquad\text{and}\qquad
\bigl[\hat \chi(0),\ \hat p	\bigr]=-\hbar \lambda\hat \chi(0),
\eq
which, together with the Eq.~\eqref{001Eq.HPZ.1.8second} and Eq.~\eqref{001Eq.HPZ.1.9}, gives the explicit form of  $\tilde{\mathbb L}_t$ reported in Eq.~\eqref{001.Eq.Master} of the main text, where \begin{subequations}\label{001Eq.HPZ.1.11}
\begin{align}
\tilde K_{11}(t)&=-\dfrac{2}{\hbar^2}\left[\dot\phi_2(t)-4\phi_2(t)\Gamma^{\text{\tiny A}}(t)+M\Delta^{\text{\tiny A}}(t)\phi_3(t)\right],
\\
\tilde K_{22}(t)&=0,
\\
\tilde K_{12}(t)&=\dfrac{1}{\hbar^2}\left[\dot\phi_3(t)-\dfrac{2}{M}\phi_2(t)+2M\Delta^{\text{\tiny A}}(t)\phi_1(t)+\right.\\
&\left.-2\phi_3(t)\Gamma^{\text{\tiny A}}(t)\right]-\dfrac{i}{\hbar}\Gamma^{\text{\tiny A}}(t).
\end{align}
\end{subequations}

\section{Explicit expression for $\Lambda^\text{\tiny dif}(t)$ and $E(t)$}
\label{app:d}

Following the procedure described in the main text, we can derive the solutions for the quadratic combinations of the position and momentum operators. Starting from Eq.~\eqref{adjoint.master}, one applies $\mathbb L_t$ to the unitary evolved operator $\hat O(t)$ written in terms of $\hat x$ and $\hat p$. Then, one applies in Eq.~\eqref{adjoint.master} the commutation relations between the operators at time $t=0$ and finds $\mathbb L_t[\hat O_t]$ depending only from operators at time $t=0$. For example, in the case of $\hat x^2$ this reads:
\bq
\begin{aligned}
&\mathbb L_t[{\hat x^2_t}]=2\dot G_1(t)\dot G_2(t){\hat x^2}+2\dot
G_1(t)\dot G_2(t){\hat p^2}/M^2+\\
&+(G_1(t)\dot G_2(t)+\dot G_1(t)G_2(t)){\{\hat x,\hat p\}}/M-2\dot\phi_1(t).
\end{aligned}
\eq
Then, one integrates the obtained expression and finds the evolution of $\hat O_t$ under the reduced dynamics. In the case of the quadratic combinations of the position and momentum operators
 the solutions are:
\bqali\label{001.Eq.gen.sol2}
\hat x^2_t&=G_1^2(t)\hat x^2+\dfrac{G_1(t)G_2(t)}{M}\{\hat x,\hat p\}+\dfrac{1}{M^2}G_2^2(t)\hat p^2+\\&-2\phi_1(t),\\
\braket{\{\hat x,\hat p\}_t}&=2M \dot G_1(t)\dot G_2(t)\braket{\hat x^2}+\dfrac{2}{M}G_1(t)G_2(t)\braket{\hat p^2}+\\&+\left(	G_1(t)\dot G_2(t)+\dot G_1(t)G_2(t)\right)\braket{\{\hat x,\hat p\}}-2\phi_3(t),\\
\braket{\hat p^2_t}&=M^2\dot G_1^2(t)\braket{\hat x^2}+M\dot G_1(t)\dot G_2(t)\braket{\{\hat x,\hat p\}}+\\&+\dot G_2^2(t)\braket{\hat p^2}-2\phi_2(t).
\eqali
Then we can compute how the system diffuses in space $\Lambda^\text{\tiny dif}(t)=\braket{\hat x_t^2}-\braket{\hat x_t}^2$:
\bqali\label{001Eq.G.6.diffus}
\Lambda^\text{\tiny dif}(t)=G_1^2(t)\left(\braket{\hat x^2}-\braket{\hat x}^2\right)+\dfrac{G_2^2(t)}{M^2}\left(\braket{\hat p^2}-\braket{\hat p}^2\right)+\\+
\dfrac{2G_1(t)G_2(t)}{M}\left(\dfrac{\braket{\left\{\hat p,\hat x\right\}}}{2}-\braket{\hat p}\braket{\hat x}\right)-2\phi_1(t).
\eqali
The energy $E(t)=\braket{\hat p^2_t}/2M+\tfrac{1}{2}M\omega_\text{\tiny S}^2\braket{\hat x_t^2}$ of the system $S$ is:
\bqali\label{001Eq.G.6.energy}
E(t)&=\dfrac{M}{2}(\omega_\text{\tiny S}^2G_1^2(t)+\dot G_1^2(t))\braket{\hat x^2}+\\&+\dfrac{(\omega_\text{\tiny S}^2G_2^2(t)+\dot G_2^2(t))}{2M}\braket{\hat p^2}+\\
&+\dfrac{1}{2}(\omega_\text{\tiny S}^2G_1(t)G_2(t)+\dot G_1(t)\dot G_2(t))\braket{\{\hat x,\hat p\}}+\\&-\left(\dfrac{\phi_2(t)}{M}+M\omega_\text{\tiny S}^2\phi_1(t)\right).
\eqali

\hfill


%

\end{document}